\documentclass[prd,aps,nofootinbib,11pt]{revtex4}
\usepackage{amsmath,amssymb,graphicx,epsfig}

\usepackage{graphics}
\usepackage{rotating}
\usepackage{epsfig}
\usepackage{color}

\newcommand{\dd}{{\rm d}}

\newcommand{\para}{\parallel}

\newcommand{\mlamB}{m_{\Lambda_b}}
\newcommand{\mlam}{m_{\Lambda}}

\begin{document}

\title{Implications of the $R_K$ and $R_{K^*}$ anomalies  }
\author{
 Wei Wang and Shuai Zhao~\footnote{shuai.zhao@sjtu.edu.cn}
 }
\affiliation{
INPAC, Shanghai Key Laboratory for Particle Physics and Cosmology, School of Physics and Astronomy, Shanghai Jiao-Tong University, Shanghai, 200240,   China}

\begin{abstract}
We discuss the implications of  the recently reported $R_K$ and
$R_{K^*}$ anomalies, the lepton flavor non-universality   in the
$B\to K\ell^+\ell^-$ and $B\to K^*\ell^+\ell^-$.  Using two sets of  hadronic inputs of form factors, we perform a fit of the new physics  to the  $R_K$ and $R_{K^*}$ data, and significant  new physics contributions  are found.  We suggest to study  the lepton flavor universality in  a number of related rare $B, B_s, B_c$ and $\Lambda_b$ decay channels,  and in particular we give the predictions for  the $\mu$-to-$e$ ratios of decay widths  with different polarizations of  the
final state particles,  and of  the $b\to d\ell^+\ell^-$ processes which
are presumably  more sensitive to the structure of the underlying new
physics. With the new physics contributions embedded in   Wilson
coefficients, we present theoretical  predictions for   lepton flavor
non-universality in these processes.
\end{abstract}

\maketitle

%%%%%%%%%%%%%%%%%%%%%%%%%%%%%%%%%%%%%%%%%%%%%%%%%%%%%%%%%%%%%%%%%%%%
%\section{Introduction}
%%%%%%%%%%%%%%%%%%%%%%%%%%%%%%%%%%%%%%%%%%%%%%%%%%%%%%%%%%%%%%%%%%%%
\section{Introduction}

The standard model (SM) of particle physics is now completed by the
discovery of   Higgs boson. Thus the  focus  in
particle physics has been gradually switched to the search for new physics
(NP) beyond the SM. This  can proceed in two distinct  ways. One is
the direct search at the high energy frontier, in which   new
particles beyond the SM are produced and detected directly. The
other is called indirect search, which is at   the high intensity
frontier. The new   particles will presumably manifest themselves as
intermediate loop effects, and might be detectable  by low-energy experiments with high precision.

In flavor physics, the  $b \to s \ell^+\ell^-$ process is a flavor
changing neutral current (FCNC) transition. This   process is  of
special interest   since  it is  induced by loop effects in the SM,
which leads to    tiny branching fractions.  Many extensions of the
SM  can generate sizable  effects that can be experimentally
validated. In particular, the   $B \to K^{*}(\to K\pi)  \mu^+ \mu^-$
decay  offers a large number of observables to test the SM, ranging
from the differential decay widths, polarizations, to  a full
analysis of angular distributions of the final state particles, for
an incomplete list one can refer to
Refs.~\cite{Hiller:2003js,Altmannshofer:2008dz,Alok:2010zd,DescotesGenon:2011yn,Altmannshofer:2011gn,Hambrock:2012dg,DescotesGenon:2012zf,Descotes-Genon:2013wba,Altmannshofer:2013foa,Buras:2013qja,Datta:2013kja,Ghosh:2014awa,Gripaios:2015gra,Barbieri:2015yvd,Feruglio:2016gvd,Hiller:2014ula,Das:2014sra,Wang:2016dne,Hu:2016gpe,Lyon:2013gba,Lyon:2014hpa}
and many references therein.

In the past few years,  quite a few observables in the channels mediated by
 $b\to s\ell^+\ell^-$ transition  have exhibited deviations from the
SM expectations.  The LHCb experiment has first  observed the
so-called $P_5^\prime$ anomaly, a sizeable discrepancy at 3.7
$\sigma$  between the measurement and the SM prediction in  one bin
for the angular observable $P_5^\prime$~\cite{Aaij:2013qta}. This
discrepancy was reproduced in a later LHCb analysis for  the  two
adjacent bins at large $K^*$ recoil~\cite{Aaij:2015oid}. To
accommodate this discrepancy,  considerable  attentions have been
paid to explore    new physics contributions (see
Refs.~\cite{Becirevic:2012dp,Chen:2016dip,Gong:2013sh,Jager:2012uw,Kou:2013gna,Sahoo:2015wya,Mahmoudi:2014mja,Ahmed:2017vsr}
and references therein), while  at the same time, this  has also
triggered the thoughts to revisit the hadronic
uncertainties~\cite{Chobanova:2017ghn,Capdevila:2017ert}.

More strikingly, the LHCb measurement  of  the
ratio~\cite{Aaij:2014ora}:
\begin{eqnarray}\label{eq:RKq2}
R_K[q_{\rm min}^2,q_{\rm max}^2]\equiv \frac{\int_{q_{\rm min}^2}^{q_{\rm max}^2}dq^2 d\Gamma(B^+ \to K^+ \mu^+ \mu^-)/dq^2}{\int_{q_{\rm min}^2}^{q_{\rm max}^2}dq^2 d\Gamma(B^+ \to K^+ e^+ e^-)/dq^2},
\end{eqnarray}
gives a hint for  the  lepton flavour universality  violation (LFUV). A plausible speculation is that
deviations from the SM are  present in $b\to s\mu^+\mu^-$ transitions instead  in $b\to s e^+e^-$ ones. Very recently the LHCb collaboration has found sizable differences between  $B\to K^*e^+e^-$ and  $B\to K^*\mu^+\mu^-$  at both low $q^2$ region and central $q^2$ region~\cite{Aaij:2017vbb}. Results for ratios
\begin{eqnarray}
\label{eq:RKstar}
R_{K^*}[q_{\rm min}^2,q_{\rm max}^2]\equiv \frac{\int_{q_{\rm min}^2}^{q_{\rm max}^2}dq^2 d\Gamma(B \to K^{*} \mu^+ \mu^-)/dq^2}{\int_{q_{\rm min}^2}^{q_{\rm max}^2}dq^2 d\Gamma(B \to K^{*} e^+ e^-)/dq^2},
\end{eqnarray}
are   given in Tab.~\ref{exp-data}, from which we can see the data
showed significant deviations from unity. These  interesting results
have  subsequently  attracted many   theoretical
attentions~\cite{Ciuchini:2017mik,Capdevila:2017bsm,Altmannshofer:2017yso,Geng:2017svp,DAmico:2017mtc,Hiller:2017bzc,Celis:2017doq,Becirevic:2017jtw,Ghosh:2017ber,Cai:2017wry,Kamenik:2017tnu,Sala:2017ihs,DiChiara:2017cjq,Alok:2017jaf,Alok:2017sui,Ellis:2017nrp,Bishara:2017pje,Tang:2017gkz,Datta:2017ezo,Das:2017kfo,Bardhan:2017xcc,Matsuzaki:2017bpp,Chiang:2017hlj,Khalil:2017mvb}.

%%%%%%%%%%%%%%%%%%%%%%%%%%%%%%%%%%%%%%%%%%%%%%%%%%%%%%%%%%%%%%%%%%%%%%%%%%%%%%%
\begin{table}[ht]
\begin{center}
\caption{Ratios of decay widths with a pair of muons and electrons in  $B\to K\ell^+\ell^-$ and $B\to K^*\ell^+\ell^-$.
}\label{exp-data}
\begin{tabular}{|c|c|c|c|c|}
\hline
Observable & SM  results   & Experimental data   \\
\hline
$R_K : q^2 = [1,6]  \text{GeV}^2$  & $1.00 \pm 0.01 $~\cite{Descotes-Genon:2015uva} &  $0.745^{+0.090}_{-0.074} \pm 0.036$~\cite{Aaij:2014ora} \\
\hline
$R_{K^*} ^{\rm low}: q^2 = [0.045,1.1] \, \text{GeV}^2$  & $0.920^{+0.007}_{-0.006} $~\cite{Geng:2017svp} &  $0.66^{+0.11}_{-0.07} \pm 0.03$~\cite{Aaij:2017vbb} \\
\hline
$R_{K^*}^{\rm central} : q^2 = [1.1,6] \, \text{GeV}^2$  & $0.996 \pm 0.002 $~\cite{Geng:2017svp} &  $0.69^{+0.11}_{-0.07} \pm 0.05$~\cite{Aaij:2017vbb} \\
\hline
\end{tabular}
\end{center}
\end{table}
%%%%%%%%%%%%%%%%%%%%%%%%%%%%%%%%%%%%%%%%%%%%%%%%%%%%%%%%%%%%%%%%%%%%%%%%%%%%%%%

The statistics significance in the data  is low at this stage, about
$3\sigma$ level. In order to obtain more conclusive results, one
should  measure the muon-versus-electron ratios  in the $B\to
K\ell^+\ell^-$ and $B\to K^*\ell^+\ell^-$ more precisely, meanwhile
one should also   investigate more channels with better
sensitivities to the structures of new physics contributions. In
this paper, we will focus on the latter.  To do so,   we will first
discuss the implications of the $R_K$ and $R_{K^*}$ anomalies in a
model-independent way, where the new particle contributions are
parameterized  in terms of  effective operators. Since there is lack
of enough data, we analyze their impact on the Wilson coefficients
of   SM operators $O_{9,10}$.  We  then propose to study the  lepton flavor
universality   in  a number of  rare $B, B_s, B_c$ and $\Lambda_b$
decay channels. Incorporating    the  new physics contributions,  we
will present  the predictions for the muon-versus-electron ratios in these  channels, making use of   various updates of form
factors~\cite{Li:2008tk,Wang:2008xt,Li:2009tx,Wang:2010ni,Horgan:2013hoa,Detmold:2012vy}.
We will demonstrate  that  the  measurements  of  lepton flavor
non-universality with different polarizations of  the final state
hadron,  and in the $b\to d\ell^+\ell^-$ processes are of great
value to decode the structure of the underlying new physics.

The rest of this paper is organized as follows. In the next section, we will use a model-independent approach and quantify  the new physics effects in terms of the short-distance Wilson coefficients. In Section  III, we will study the LFUV in various FCNC channels.
Our conclusion  is given in the last section.

%%%%%%%%%%%%%%%%%%%%%%%%%%%%%%%%%%%%%%%%%%%
\section{Implications from  the $R_K$ and $R_{K^*}$}
%%%%%%%%%%%%%%%%%%%%%%%%%%%%%%%%%%%%%%%%%%%

In this section, we will first  study the impact of the  $R_K$ and $R_{K^*}$   data.
In the SM, the effective Hamiltonian for the transition  $b\to s\ell^+\ell^-$
 \begin{eqnarray}
 {\cal
 H}_{\rm{eff}}=
 -\frac{G_F}{\sqrt{2}}V_{tb}V^*_{ts}\sum_{i=1}^{10}C_i(\mu)O_i(\mu)
 \nonumber\label{eq:Hamiltonian}
 \end{eqnarray}
involves the  four-quark and the magnetic penguin operators $O_i$. Here  $C_i(\mu)$
are the    Wilson coefficients for these local operators $O_i$.
$G_F$ is the Fermi constant, $V_{tb}$ and
$V_{ts}$ are   CKM matrix elements.   The dominant contributions to $b\to s\ell^+\ell^-$  come from the following operators:
\begin{eqnarray}
 %-------------------------------------------------
 O_7&=&\frac{e m_b}{8\pi^2}\bar
 s\sigma^{\mu\nu}(1+\gamma_5)bF_{\mu\nu}+\frac{e m_s}{8\pi^2}\bar
 s\sigma^{\mu\nu}(1-\gamma_5)bF_{\mu\nu},\nonumber\\
 %------------------------------------------------------
 O_9&=&\frac{\alpha_{\rm{em}}}{2\pi}(\bar l\gamma_{\mu}l) \bar
 s\gamma^{\mu}(1-\gamma_5)b ,\;\;
 O_{10}=\frac{\alpha_{\rm{em}}}{2\pi}(\bar l\gamma_{\mu}\gamma_5l) \bar
 s\gamma^{\mu}(1-\gamma_5)b.\label{eq:operators}
 \end{eqnarray}

 The above effective
Hamiltonian gives  the $B\to K\ell^+\ell^-$ decay width  as:
\begin{eqnarray}
 \frac{d\Gamma(B\to K\ell^+\ell^-)}{dq^2}  &=& \frac{ G_F^2  \sqrt \lambda \alpha_{em}^2\beta_l}{1536m_B^3\pi^5}  |V_{tb}V_{ts}^*|^2
\times \left[ \lambda (1+2\hat m_l^2)   \left| {C_9}
   f_+(q^2)    +C_{7}   \frac{2m_b f_T(q^2)}{ m_B+m_K }\right|^2\right. \nonumber\\
 &&\left.+  \lambda \beta_l^2 \left|  C_{10}  \right|^2 f_+^2 (q^2)  +    6\hat m_l^2 \left|
 C_{10}\right|^2(m_B^2-m_K^2)^2 f_0^2(q^2)\right],
\end{eqnarray}
where $\hat m_l=m_l/\sqrt{q^2}$,  $\beta_l=\sqrt{1-\hat m_l^2}$,
$\lambda= (m^2_{B}-m^2_{K}-q^2)^2-4m^2_{K}q^2$, and $f_+$,  $f_0$
and $f_T$ are the $B\to K$ form factors. In the above
expression, we have neglected the non-factorizable contributions
which are expected to be negligible for $R_K$.

The    decay width for $B\to K^*\ell^+\ell^-$ can be  derived  in
terms of the helicity
amplitude~\cite{Lu:2011jm,Doring:2013wka,Dey:2015rqa,Gratrex:2015hna,Dey:2016oun}.
The differential decay width is given as
\begin{eqnarray}
 \frac{d\Gamma(B\to K^* \ell^+ \ell^-)}{dq^2}
 &=& \frac{3}{4}\Big(I_1^c  + 2I_1^s\Big)-\frac14
 \Big(I_2^c+2I_2^s\Big),
\end{eqnarray}
with
\begin{eqnarray}\label{eq:I1I2}
 I_1^c&=&  (|A^1_{L0}|^2+|A^1_{R0}|^2)
 +8  \hat m_l^2 {\rm Re}[A^1_{L0}A^{1*}_{R0} ]+4 \hat m_l^2  |A^1_t|^2, \nonumber\\
 I_1^s&=& \left(3/4- \hat m_l^2 \right)[|A^1_{L\perp}|^2
 +|A^1_{L||}|^2+|A^1_{R\perp}|^2+|A^1_{R||}|^2  ]
+ 4 \hat m_l^2  {\rm Re}[A^1_{L\perp}A^{1*}_{R\perp}
 + A^1_{L||}A^{1*}_{R||}],\nonumber\\
 I_2^c  &=& -\beta_l^2(  |A^1_{L0}|^2+ |A^1_{R0}|^2),\nonumber\\
 I_2^s  &=&
 \frac{1}{4}\beta_l^2(|A^1_{L\perp}|^2+|A^1_{L||}|^2+|A^1_{R\perp}|^2+|A^1_{R||}|^2). \label{eq:angularCoefficients}
\end{eqnarray}
The handedness  label $L$ or $R$ corresponds to  the chirality of the
di-lepton system. Functions $A_{L/Ri}$  can be  expressed in terms of $B\to
K^*$ form factors
\begin{eqnarray}
A^1_t&=& 2 \sqrt{N_{K_J^*}} N_{1}  C_{10}\frac{\sqrt \lambda}{\sqrt
{q^2}}A_0(q^2),\\
  A^1_{L 0 }&=&\frac{   N_1 \sqrt{ N_{K_J^*}}  }{2m_{K^*_J}\sqrt {q^2}}\left[ (C_9-C_{10})  [(m_B^2-m_{K^*}^2-q^2)(m_B+m_{K^*})A_1
 -\frac{\lambda}{m_B+m_{K^*}}A_2]\right.\nonumber\\
 &&\left. + 2m_b C_7  [ (m_B^2+3m_{K^*}^2-q^2)T_2 -\frac{\lambda  }{m_B^2-m_{K^*}^2}T_3]\right],   \\
 A^1_{L \perp} &=& -   \sqrt{2N_{K_J^*}}
N_1\Big[(C_9-C_{10})
 \frac{\sqrt \lambda V}{m_B+m_{K^*}} +\frac{2m_b C_7}{q^2}\sqrt \lambda T_1\Big], \label{eq:A1Lperp}\\
 A^1_{L ||}&=&   \sqrt{2N_{K_J^*}} N_{1}
\Big[(C_9-C_{10}) (m_B+m_{K^*})A_1 +\frac{2m_b C_{7}
}{q^2}(m_B^2-m_{K^*}^2)
 T_2 \Big]\,,\label{eq:A1Lpar}
 \end{eqnarray}
 with $N_1= \frac{iG_F}{4\sqrt
2} \frac{\alpha_{\rm em}}{\pi} V_{tb}V_{ts}^*$, $ N_{K_J^*} =
{8/{3}} {\sqrt {\lambda} {q^2}\beta_l}/({256\pi^3 m_B^3})$ and
$\lambda\equiv (m^2_{B}-m^2_{K^*}-q^2)^2-4m^2_{K^*}q^2$.
 The right-handed decay amplitudes are  obtained by reversing the sign of $C_{10}$:
\begin{eqnarray}
 A_{Ri}
  &=& A_{Li}|_{C_{10}\to -C_{10}}.
\end{eqnarray}

Within the SM, one can easily  find that
 results for $R_K$ and $R_{K^*}$ are   extremely close to 1 and thus deviate
from the experimental data. If new physics  is indeed present, it can be in $b\to s\mu^+\mu^-$ and/or  $b\to se^+e^-$transitions. In order to explain the  $R_K$ and $R_{K^*}$ data, one can
enhance the partial width for the electronic mode or reduce the one
for the muonic mode.  It seems that the SM result for the $B\to Ke^+e^-$ is consistent with the data, and thus here we will adopt the strategy  that the
muonic decay width is reduced by new physics.

After integrating out the high scale intermediate states the new
physics contributions can   be incorporated into the effective
operators. As there is lack of enough data that shows significant deviations  with SM, we will assume that   NP contributions
can be incorporated into   Wilson coefficients $C_{9}$ and $C_{10}$.
For this purpose, we define
\begin{eqnarray}
\delta C_{9}^\mu =C_{9}^\mu -C_9^{\rm SM},\;\;
\delta C_{10}^\mu =C_{10}^\mu -C_{10}^{\rm SM}.
\end{eqnarray}
The   $O_{7}$ contribution to $b\to s\ell^+\ell^-$   arises from the
coupling of  a photon with the lepton pair. On one hand, this
coupling  is  highly constrained from the $b\to s\gamma$ data. On the other
hand, this coefficient is flavor blinded and thus even if NP affect
$C_7$, the $\mu$-to-$e$ will not be affected.

\begin{figure}\begin{center}
\includegraphics[scale=0.32]{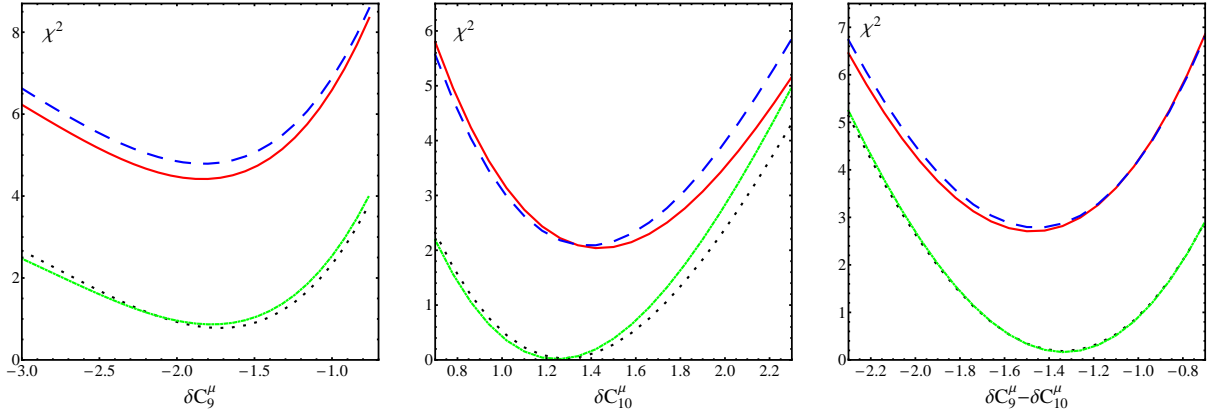}
\caption{Impact of $R_K$ and $R_{K^*}$ data on the  $\delta
C_9^{\mu}$ (left panel), $\delta C_{10}^{\mu}$ (central panel) or
$\delta C_9^{\mu}-\delta C_{10}^{\mu}$ (right panel).   The
dependence of the total $\chi^2$   for all  data in Tab. I on
Wilson coefficients is shown as the solid (red) and dashed (blue)
curves, which correspond to the form factors from
LQCD~\cite{Bouchard:2013pna,Horgan:2013hoa} and
LCSR~\cite{Ball:2004ye,Ball:2004rg}, respectively.  Removing the
low-$q^2$ data for  $B\to K^*\ell^+\ell^-$, the results are shown as
dotted (black) and  and dot-dashed  (green) curves.  }
\label{fig:fitted_results}
\end{center}
\end{figure}

For the analysis, we adopt three scenarios,
\begin{enumerate}
\item Only $C_{9}$ is affected  with $\delta C_{9}^\mu\neq0$.
\item Only $C_{10}$ is affected  with $\delta C_{10}^\mu\neq0$.
\item Both  $C_{9}$ and $C_{10}$ are   affected in the form:  $\delta C_{9}^\mu=-\delta C_{10}^\mu\neq0$.
\end{enumerate}
Using the $R_K$ and $R_{K^*}$ data, we show our results in
FIG.~\ref{fig:fitted_results}. The left panel corresponds to
scenario 1, and  the middle panel corresponds to the constraint on
$\delta C_{10}^\mu$, the last one corresponds to the scenario 3 with
a nonzero $\delta C^{\mu}_{9}- \delta C^\mu_{10}$. In this analysis,
we have used  two sets of  $B\to K$ and $B\to K^*$ form factors. One
is from the light-cone sum rules
(LCSR)~\cite{Ball:2004ye,Ball:2004rg,Straub:2015ica}, corresponding
to the dashed curves. The other   is from Lattice QCD
(LQCD)~\cite{Bouchard:2013pna,Horgan:2013hoa}, which gives the solid
curves. As one can see clearly from the figure,  the results are not
sensitive to the form factors, and this also  partly validate the
neglect of other hadronic uncertainties like non-factorizable
contributions.  Using the LQCD set of form
factors~\cite{Bouchard:2013pna,Horgan:2013hoa} and the data in
Tab.I,  we found the best-fitted  central value and the $1\sigma$
range for $\delta C_{9}^\mu$  in scenario 1 as
\begin{eqnarray}
\delta C_{9}^\mu = -1.83\,, \;\;\;  -2.63< \delta C_{9}^\mu<
-1.25\,. \label{eq:C9_Constrained}
\end{eqnarray}
For scenario 2, we have
\begin{eqnarray}
\delta C_{10}^\mu = 1.43, \;\;\;  1.04< \delta C_{10}^\mu< 1.89,
\end{eqnarray}
while  for the $\delta C_{9}^\mu =- \delta C_{10}^\mu$, we obtain
\begin{eqnarray}
\delta C_{9}^\mu  - \delta C_{10}^\mu  = -1.47, \;\;\;  -1.89<
\delta C_{9}^\mu- \delta C_{10}^\mu    < -1.08.
\end{eqnarray}

%%%%%%%%%%%%%%%%%%%%%%%%%%%%%%%%%%%%%%%%%%%%%%%%%%%%%%%%%%%%%%%%%%%%%%%%%%%%%%
%%%%%%%%%%%%%%%%%%%%%%%%%%%%%%%%%%%%%%%%%%%%%%%%%%%%%%%%%%%%%%%%%%%%%%%%%%%%%%%%

A few  remarks are given in order.
\begin{figure}
\begin{center}
\includegraphics[width=1\textwidth]{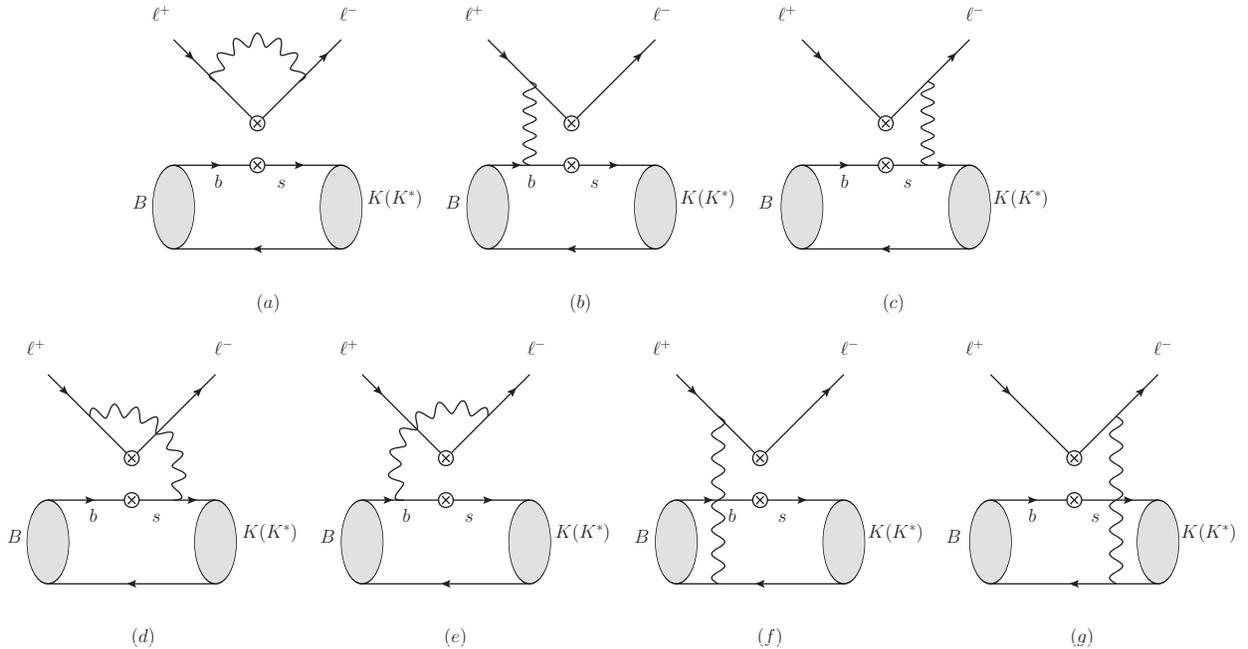}
\caption{The electromagnetic corrections to $B\to K\ell^-\ell^+$ and
$B\to K^*\ell^-\ell^+.$ } \label{fig:BtoK}
\end{center}
\end{figure}
\begin{itemize}

\item Since the Wilson coefficient in the electron channel is unchanged,  the $\delta C_9^\mu$ and $\delta C_{10}^\mu$ could be viewed as the difference between the  Wilson coefficients for the lepton and muon case.

\item
We have found the largest deviation between the fitted results and the data comes from  the low-$q^2$ region.   Removing this data, we show the $\chi^2$ in FIG.~\ref{fig:fitted_results} as dotted and dot-dashed curves, where the $\chi^2$ has   been greatly reduced.  The reason is that in low-$q^2$ region, the dominant contribution to $R_{K^*}$ arises from the transverse polarization of $K^*$. From Eq.~\eqref{eq:A1Lperp} and ~\eqref{eq:A1Lpar}, one can see this contribution is dominated by  $O_7$ and  less sensitive to $O_{9,10}$.
A light mediator that only couples to the $\mu^+\mu^-$ is explored for instance in Refs.~\cite{Sala:2017ihs,Bishara:2017pje,Datta:2017ezo}.

\item For the  $R_K$ and $R_{K^*}$ predictions  in Refs.~\cite{Descotes-Genon:2015uva,Geng:2017svp},
theoretical errors are typically less than one percent, while
Ref.~\cite{Bobeth:2007dw}  gives the prediction with even smaller
uncertainty $R_K= 1.0003\pm 0.0001$. However it is necessary to
stress that  these  results did not consider the electromagnetic
corrections properly. We give the Feynman diagrams in
Fig.~\ref{fig:BtoK}. Fig.~\ref{fig:BtoK}(a) is the typical Sudakov
form factor, which usually introduces a double logarithm in terms of
$\alpha/\pi \ln (q^2/m_\ell^2)$.  The difference between the  double
logarithms for the electron and muon mode is about $3\%$.  A complete analysis requests the detailed calculation of all diagrams
in  Fig.~\ref{fig:BtoK} and analyses can be found  in Ref.~\cite{Bordone:2016gaq}.   The nonfactorizable corrections to the
amplitude can be found in Ref.~\cite{Beneke:2001at}.

\item It is necessary to point out that   there are a number of observables in $B\to K \mu^+\mu^-$ and $B\to K^* \mu^+\mu^-$ that have been experimentally measured.
These observables are of great values to provide very stringent
constraints on the Wilson coefficients in the factorization
approach. On the other hand, most of these observables in $B\to K
\mu^+\mu^-$ and $B\to K^* \mu^+\mu^-$ are not sensitive to the
flavor non-universality coupling since only the mu lepton is
involved. The exploration of the $\mu$-to-$e$ ratios will be able to
detect the  difference in the new physics couplings to fermions.
 It is always meaningful to conduct  a comprehensive global analysis and  incorporate as many observables as possible.
  At this stage, the study of flavor non-universality in flavor physics is at the beginning, and we believe measuring more
  $\mu$ to $e$ ratios (for instance the ones in Table II shown in the following section) will be helpful.

\item For a more comprehensive analysis, one may combine various experimental data on the flavor changing neutral current processes for instance  in Refs.~\cite{Ciuchini:2017mik,Capdevila:2017bsm,Altmannshofer:2017yso,Geng:2017svp,DAmico:2017mtc}. We quote the results in scenario I in Ref.~\cite{Ciuchini:2017mik},
\begin{eqnarray}
 \delta C_{9}^{\mu} =-1.58\pm0.28, \;\;\;  \delta C_{9}^{e} =-0.10\pm0.45,
\end{eqnarray}
from which we can see that the results are close to our scenario 1.  This implies that for the determination of flavor dependent Wilson coefficient, the $R_K$ and $R_{K^*}$ are dominant.  From a practical viewpoint,   since the main purpose of this paper is to explore the implications of the large lepton flavor non-universality, we will use our fitted results  to predict the lepton flavor non-universality for a number of other channels.

\end{itemize}

\begin{figure}\begin{center}
\includegraphics[scale=0.6]{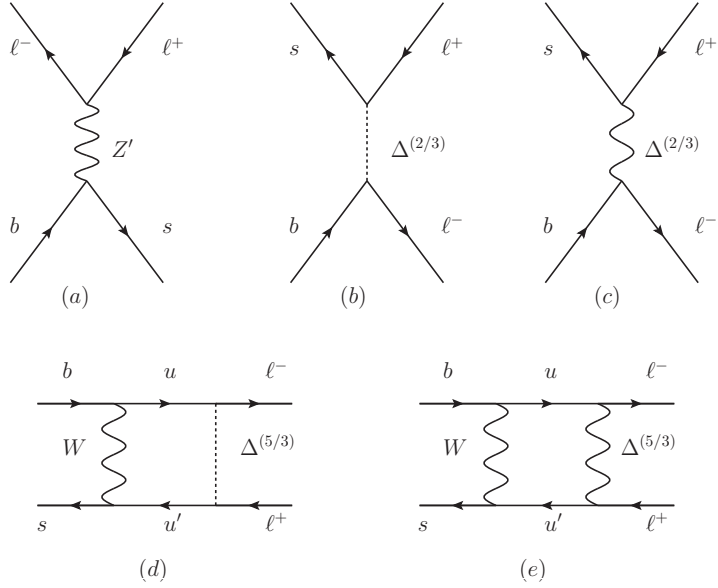}
\caption{New physics scenarios  that can contribute  to $b\to
s\mu^+\mu^-$. The panel (a) shows a $Z'$, and in  the other four
diagrams  $\Delta$ denotes a leptoquark with different spins and
charges.   } \label{fig:NP}
\end{center}
\end{figure}

Explicit models which can realize these scenarios  include the
flavor non-universal $Z'$ model, leptoquark model and vector-like
models, see, e.g.,
Refs.~\cite{Chang:2011jka,Gauld:2013qba,Chang:2013hba,Hiller:2014yaa,Gripaios:2014tna,Calibbi:2015kma,Alonso:2015sja,Becirevic:2015asa,Cox:2016epl,Becirevic:2016yqi,Greljo:2015mma,Li:2016vvp,Hu:2017qxj,Chen:2017hir,Crivellin:2015mga,Crivellin:2015lwa,Sierra:2015fma,Crivellin:2015era,Becirevic:2016zri,Celis:2015ara,Falkowski:2015zwa,Allanach:2015gkd,Fuyuto:2015gmk,Chiang:2016qov,Crivellin:2016ejn,Datta:2017pfz,Megias:2016bde,Megias:2017ove,Belanger:2015nma,Carmona:2015ena}
and many references therein. Their generic  contributions are shown
in FIG.~\ref{fig:NP}. Taking the $Z'$ model as an example,  the SM
can be extended by including an additional $U(1)^{\prime}$ symmetry,
which can leads to the Lagrangian of $Z^{\prime}\bar bs$ couplings
\begin{eqnarray}
{\cal L}_{\rm{FCNC}}^{Z^{\prime}}=-g^{\prime}(B_{sb}^L\bar
s_L\gamma_{\mu}b_L + B_{sb}^R\bar s_R\gamma_{\mu}b_R)Z^{\prime\mu} +
{\rm h.c.}\, . \label{eq:HamiltonZ}
\end{eqnarray}
It contributes to the $b\to s\ell^+\ell^-$ decay at tree level
\begin{eqnarray}
 {\cal H}_{\rm{eff}}^{Z^{\prime}}=
\frac{8G_F}{\sqrt{2}} (\rho_{sb}^L\bar s_L\gamma_{\mu}b_L +
\rho_{sb}^R\bar s_R\gamma_{\mu}b_R) (\rho_{ll}^L\bar
\ell_L\gamma^{\mu}\ell_L +\rho_{ll}^R\bar \ell_R\gamma^{\mu}\ell_R)
~, \label{eq:HamiltonZprime}
\end{eqnarray}
where the coupling is
\begin{eqnarray}
\rho_{ff'}^{L,R} \equiv \frac{g'M_Z}{gM_{Z'}} B_{ff'}^{L,R}
\end{eqnarray}
where  the   $g$ standard model $SU(2)_L$ coupling.  For simplicity, one  can  assume that  the FCNC
couplings of the $Z^{\prime}$ and quarks only occur in the
left-handed sector:  $\rho_{sb}^R=0$. Thus in this case the effects of
the $Z^{\prime}$  will  modify the Wilson coefficients
$C_9$ and $C_{10}$:
\begin{eqnarray}
  C_9^{Z^{\prime}} &=& C_9^{} -\frac{4\pi}{\alpha_{\rm em}}  {\frac{  \rho_{sb}^L
(\rho_{ll}^L+\rho_{ll}^R)}{
 V_{tb}V^*_{ts} }},\;\;\;
  C_{10}^{Z^{\prime}} =C_{10}+\frac{4\pi}{\alpha_{\rm em}}{  \frac{ \rho_{sb}^L(\rho_{ll}^L-\rho_{ll}^R) }{
  V_{tb}V^*_{ts} }}.
  \label{eq:wilsonZprime}
\end{eqnarray}
From this expression, we can see that the $\delta C_{9}^\mu$ and $\delta C_{10}^\mu$ are not entirely correlated.   This corresponds to the scenario 1 and 2 in our previous analysis.

The impact in a leptoquark model has  been discussed for instance in Ref.~\cite{Becirevic:2017jtw}, where the NP contribution satisfies
\begin{eqnarray}
 \delta C_{9}^{LQ,\mu}=-\delta C_{10}^{LQ,\mu}.
\end{eqnarray}
This corresponds to the scenario 3.

%A leptoquark can be a scalar or a vector with different charges. A
%scalar leptoquark as shown in FIG.~(\ref{fig:NP}b) can leads to a
%low-energy effective operator
%\begin{eqnarray}
% O\sim [\bar \ell b] [\bar s\ell].
%\end{eqnarray}
%After performing the Fierz transformation, the low energy operators
%will have structures that differ with $O_9$ and $O_{10}$.

%%%%%%%%%%%%%%%%%%%%%%%%%%%%%%%%%%%%%%%%%%%%%%%%%%%%%%%%%%%%%%%%%%%%%%%%%%%%%%
%%%%%%%%%%%%%%%%%%%%%%%%%%%%%%%%%%%%%%%%%%%%%%%%%%%%%%%%%%%%%%%%%%%%%%%%%%%%%%%%

%%%%%%%%%%%%%%%%%%%%%%%%%%
\section{Lepton Flavor Universality in   FCNC channels }
%%%%%%%%%%%%%%%%%%%%%%%%%%

In this section, we will study the $\mu$-to-$e$ ratios of decay
widths  in various FCNC  channels.  Since the three scenarios considered in the last section
describe the data equally well, we will choose the first one for
illustration in the following. We follow a similar definition
\begin{eqnarray}
\label{eq:RKq2}
R_{B, M}[q_{\rm min}^2,q_{\rm max}^2]\equiv \frac{\int_{q_{\rm min}^2}^{q_{\rm max}^2}dq^2 d\Gamma(B  \to M \mu^+ \mu^-)/dq^2}{\int_{q_{\rm min}^2}^{q_{\rm max}^2}dq^2 d\Gamma(B  \to M e^+ e^-)/dq^2},\label{eq:RRM_deff}
\end{eqnarray}
where $B$ denotes a heavy particle and $M$ denotes a final state.
The channels to be studied include  $B\to K^*_{0,2}(1430)\ell^+\ell^-$,
$B_s\to f_0(980)\ell^+\ell^-$, $B\to K_1(1270)\ell^+\ell^-$,
$B_s\to f_2(1525)\ell^+\ell^-$, $B_s\to \phi \ell^+\ell^-$, $B_c\to
D_s\ell^+\ell^-$, $B_c\to D_s^*\ell^+\ell^-$.  The expressions for
their decay widths have been given in the last section.  In
addition, we will also analyze  on the $R$ ratio for  the baryonic
decay $\Lambda_b\to \Lambda\ell^+\ell^-$. The differential decay
width for $\Lambda_b\to \Lambda\ell^+\ell^-$ is given
as~\cite{Boer:2014kda}
\begin{equation}
    \frac{\dd \Gamma}{\dd q^2}[\Lambda_b\to \Lambda\ell^+\ell^-]  = 2 K_{1ss} + K_{1cc} \,, \label{eq:Lambda_width}
\end{equation}
where
\begin{eqnarray}
    K_{1ss}(q^2) & =& \frac{1}{4}\Big[|A_{\perp_1}^R|^2 + |A_{\para_1}^R|^2
    + 2 |A_{\perp_0}^R|^2 + 2 |A_{\para_0}^R|^2 + (R \leftrightarrow L)\Big] ,\nonumber\\
    K_{1cc}(q^2) & =& \frac{1}{2}\Big[|A_{\perp_1}^R|^2 + |A_{\para_1}^R|^2 + (R \leftrightarrow L)\Big].
\end{eqnarray}
The functions $A$ are defined  as
\begin{eqnarray}\label{eq:Afunction}
    A_{\perp_1}^{L(R)} & =&  \sqrt{2} N \left[\left(C_9\mp C_{10}\right) H_+^V
    - \frac{2 m_b C_7}{q^2} H_+^{T }\right],\;\;
    A_{\para_1}^{L(R)}  = -\sqrt{2} N \left[\left(C_9\mp C_{10}\right)H_+^A
    + \frac{2 m_b  C_7}{q^2} H_+^{T5}\right],\nonumber\\
    A_{\perp_0}^{L(R)} & =&  \sqrt{2} N \left[\left(C_9\mp C_{10}\right) H_0^V
    - \frac{2 m_b  C_7}{q^2} H_0^{T }\right]\,,\;
    A_{\para_0}^{L(R)}   = -\sqrt{2} N \left[\left(C_9\mp C_{10}\right) H_0^A
    + \frac{2 m_b  C_7 }{q^2} H_0^{T5}\right],\nonumber\\ \label{eq:lambda_b_decay}
\end{eqnarray}
where the normalization factor $N$ is
\begin{eqnarray}
  N=G_F V_{tb}V^*_{ts} \alpha_{\rm em}\sqrt{\frac{q^2\sqrt{\lambda(m^2_{\Lambda_b},m^2_{\Lambda},q^2)}}{3\cdot 2^{11}
  m^3_{\Lambda_b}\pi^5}}.
\end{eqnarray}
 The helicity amplitudes are given by
\begin{eqnarray}
    H_0^V & =& f_0^V(q^2)\, \frac{m_{\Lambda_b} + m_\Lambda}{\sqrt{q^2}} \, \sqrt{s_-},\;\;\;\;\;
    H_+^V = -f_\perp^V(q^2) \, \sqrt{2 s_-}, \nonumber\\
    H_0^A  & =& f_0^A(q^2) \, \frac{m_{\Lambda_b} - m_\Lambda}{\sqrt{q^2}} \, \sqrt{s_+},\;\;\;\;\;
    H_+^A  = - f_\perp^A(q^2) \, \sqrt{2 s_+}\, \nonumber\\
    H_0^T & =& -f_0^T(q^2) \, \sqrt{q^2} \, \sqrt{s_-}\,,\;\;\;\;\;
    H_+^T =f_\perp^T(q^2) \, (\mlamB + \mlam) \,  \sqrt{2 s_-}\,, \nonumber\\
    H_0^{T5}       & =&  f_0^{T5}(q^2) \, \sqrt{q^2} \, \sqrt{s_+}\,,\;\;\;\;\;
    H_+^{T5}
        = - f_\perp^{T5}(q^2) \, (\mlamB - \mlam) \,  \sqrt{2 s_+},
\end{eqnarray}
where $s_\pm \equiv (m_{\Lambda_b} \pm m_\Lambda)^2 - q^2. $ The
$f^{i}_{0/\perp}$ with $i=V, A, T, T5$ are the  $\Lambda_b\to \Lambda$ form factors.

The $B_s\to \phi\ell^+\ell^-$ and $\Lambda_b\to \Lambda$ form
factors are used from  LQCD calculation in
Refs.~\cite{Horgan:2013hoa,Detmold:2016pkz}, respectively. The $B\to
K_0^*(1430)$ and $B_s\to f_0(980)$ form factors are taken from
Ref.~\cite{Li:2008tk,Colangelo:2010bg}. The $B\to K_1(1270)$ form
factors are calculated in the perturbative QCD
approach~\cite{Li:2009tx}, and the mixing angle between
$K_1(1^{++})$ and $K_1(1^{+-})$ is set to be approximately
$45^\circ$. In this case the $B\to K_1(1400)\ell^+\ell^-$ is greatly
suppresed~\cite{Li:2009rc}.   The $B\to K_2$ and $B_s\to f_2(1525)$
form factors are taken from Ref.~\cite{Wang:2010ni}.  The $B_c\to
D_s/D_s^*$ form factors are provided in light-front quark
model~\cite{Wang:2008xt}, and in this work we have calculated  the
previously-missing tensor form factors. Using the Wilson coefficient
$\delta C_9^\mu $ in Eq.~\eqref{eq:C9_Constrained}, we present  our
numerical results for $R_{B, M}$ in TABLE~\ref{tab:result_RM}. Three
kinematics regions are chosen in the analysis: low  $q^2$ with
[0.045, 1] ${\rm GeV}^2$, central $q^2$  with [1, 6] ${\rm GeV}^2$
and high $q^2$ region with  [14 ${\rm GeV}^2$, $q^2_{\rm
max}=(m_{B}-m_M)^2$]. For a vector final state, the longitudinal and
transverse polarizations are separated and labeled as $L$ and $T$,
respectively.  For $\Lambda_b\to \Lambda\ell^+\ell^-$, a similar
decomposition is used, in which the superscript  $0$ means the
$\Lambda_b$ and $\Lambda$ have the same polarization  while $1$
corresponds to different polarizations. The SM predictions for these
ratios are listed in Tab.~\ref{tab:result_SM}.

%%%%%%%%%%%%%%%%%%%%%%%%%%%%%%%%%%%%
\begin{table}
 \caption{Theoretical results for the $\mu$-to-$e$ ratio  $R_{B,M}$ of decay widths as defined in Eq.~\eqref{eq:RRM_deff} in various $b\to s\ell^+\ell^-$ channels. Three kinematics regions are chosen: low, central and high $q^2$ regions.   Wilson coefficient $C_9$ is used as in  Eq.~\eqref{eq:C9_Constrained} based on the analysis of $R_K$ and $R_{K^*}$. For a vector final state, the longitudinal and transverse polarizations are separated and labeled as $L$ and $T$, respectively.
  For  $\Lambda_b\to \Lambda\ell^+\ell^-$, a similar decomposition is used: the superscript
   $0$ means that the $\Lambda_b$ and  $\Lambda$ have the same polarization,
    while $1$ corresponds to different polarizations.   }
 \label{tab:result_RM}
 \begin{center}
 \begin{tabular}{|c|c|c|c|}
 \hline
  Observable               & Low $q^2: [0.045, 1] {\rm GeV}^2$                    & Central $q^2: [ 1, 6] {\rm GeV}^2$               &  High $q^2:[14{\rm GeV}^2, q^2_{\rm max} ]$ \\ \hline
  $R_{B, K_0^*(1430)}$     &  $0.688_{-0.073}^{+0.075}$  &  $0.702_{-0.075}^{+0.076}$   &  $0.721_{-0.074}^{+0.074}$\\ \hline
  %%%%%%%%%%%%%%%%%%%%%%%%%%%%%%
  $R_{B_s, f_0(980)}$    &  $0.687_{-0.074}^{+0.074}$  &  $0.700_{-0.076}^{+0.076}$    &   $0.707_{-0.074}^{+0.075}$\\ \hline
  %%%%%%%%%%%%%%%%%%%%%%%%%%%%%%
  $R_{B_c, D_s}$           &  $0.686_{-0.075}^{+0.075}$  &  $0.699_{-0.077}^{+0.077}$   & $0.706_{-0.076}^{+0.076}$\\ \hline
  %%%%%%%%%%%%%%%%%%%%%%%%%%%%%%
  $R_{B_s, \phi}$          &  $0.863^{+0.016}_{-0.010}$  &  $0.772^{+0.051}_{-0.040}$   &  $0.710_{-0.067}^{+0.071}$ \\
  $R_{B_s, \phi}^L $       &  $0.697^{+0.074}_{-0.074}$  &  $0.701^{+0.076}_{-0.076}$   & $0.706_{-0.071}^{+0.073}$\\
  $R_{B_s, \phi}^T $       &  $0.975^{-0.024}_{+0.034}$  &  $1.059^{-0.049}_{+0.108}$   & $0.712_{-0.065}^{+0.070}$ \\    \hline
  %%%%%%%%%%%%%%%%%%%%%%%%%%%%%%
  $R_{B_c, D^*_s}$         &  $0.926_{+0.012}^{-0.006}$  &  $0.940_{+0.034}^{-0.003}$   &  $0.749_{-0.041}^{+0.056}$ \\
  $R_{B_c, D^*_s}^L$       &  $0.704_{-0.059}^{+0.066}$  &  $0.719_{-0.060}^{+0.067}$   &  $0.736_{-0.049}^{+0.060}$\\
  $R_{B_c, D^*_s}^T$       &  $0.956_{+0.021}^{-0.015}$  &  $1.289_{+0.182}^{-0.113}$   &  $0.756_{-0.037}^{+0.053}$ \\     \hline
  %%%%%%%%%%%%%%%%%%%%%%%%%%%%%%
  $R_{B,   K^*_2}$         &  $0.851_{-0.011}^{+0.017}$  &
  $0.759_{-0.044}^{+0.055}$ &  $0.718_{-0.062}^{+0.068}$\\
  $R_{B,   K^*_2}^L$       &  $0.675_{-0.076}^{+0.075}$  &
  $0.696_{-0.077}^{+0.077}$  &   $0.713_{-0.065}^{+0.070}$\\
  $R_{B,   K^*_2}^T$       &  $0.983_{+0.038}^{-0.026}$  &
  $1.051_{+0.109}^{-0.049}$  &  $0.721_{-0.059}^{+0.066}$\\   \hline
  %%%%%%%%%%%%%%%%%%%%%%%%%%%%%%
  $R_{B_s,   f_2}$       &  $0.858_{-0.008}^{+0.014}$  &  $0.767_{-0.040}^{+0.052}$   &  $0.720_{-0.060}^{+0.067}$ \\
  $R_{B_s,   f_2}^L$     &  $0.675_{-0.075}^{+0.075}$  &  $0.697_{-0.076}^{+0.076}$   & $0.716_{-0.063}^{+0.069}$ \\
  $R_{B_s,   f_2}^T$     &  $0.982_{+0.037}^{-0.026}$  &  $1.063_{+0.114}^{-0.052}$   &  $0.723_{-0.058}^{+0.065}$\\ \hline
  %%%%%%%%%%%%%%%%%%%%%%%%%%%%%%
  $R_{B,   K_1(1270)}$         &  $0.909_{-0.004}^{+0.008}$  &
  $0.880_{-0.002}^{+0.002}$ &  $0.714_{-0.065}^{+0.069}$\\
  $R_{B,   K_1(1270)}^L$       &  $0.751_{-0.094}^{+0.085}$  &
  $0.717_{-0.100}^{+0.088}$  &   $0.712_{-0.067}^{+0.071}$\\
  $R_{B,   K_1(1270)}^T$       &  $0.978_{+0.036}^{-0.025}$  &
  $1.078_{+0.118}^{-0.056}$  &  $0.714_{-0.064}^{+0.069}$\\   \hline
  %%%%%%%%%%%%%%%%%%%%%%%%%%%%%%
  $R_{\Lambda_b, \Lambda}$ &  $0.931_{-0.007}^{+0.014}$  &   $0.773_{-0.039}^{+0.051}$  &  $0.712_{-0.068}^{+0.071}$ \\
  $R_{\Lambda_b, \Lambda}^0 $ & $0.708_{-0.070}^{+0.073}$ &  $0.705_{-0.072}^{+0.074}$  &  $0.707_{-0.072}^{+0.073}$\\
  $R_{\Lambda_b, \Lambda}^1 $ & $1.071_{+0.032}^{-0.023}$ &  $1.104_{+0.124}^{-0.060}$  &  $0.715_{-0.065}^{+0.070}$\\  \hline
 \end{tabular}
 \end{center}
 \end{table}
 %%%%%%%%%%%%%%%%%%%%%%%%%%%%%%%%%%%%

%%%%%%%%%%%%%%%%%%%%%%%%%%%%%%%%%%%%
\begin{table}
 \caption{Theoretical results for the $\mu$-to-$e$ ratio  $R_{B,M}$ of decay widths as defined in Eq.~\eqref{eq:RRM_deff} in various $b\to s\ell^+\ell^-$ channels in the SM.
  Three kinematics regions are chosen: low, central and high $q^2$ regions.
  For a vector final state, the longitudinal and transverse polarizations are separated and labeled as $L$ and $T$, respectively.  We do not present the results  $\Lambda_b\to \Lambda\ell^+\ell^-$ since the lepton mass effects are not included in Eq.~\eqref{eq:lambda_b_decay}.      }
 \label{tab:result_SM}
 \begin{center}
 \begin{tabular}{|c|c|c|c|}
 \hline
  Observable               & Low $q^2: [0.045, 1] {\rm GeV}^2$                    & Central $q^2: [ 1, 6] {\rm GeV}^2$               &  High $q^2:[14{\rm GeV}^2, q^2_{\rm max} ]$ \\ \hline
  $R_{B, K_0^*(1430)}$     &  $0.980$  &  $1.001$   &  $1.029$\\ \hline
  %%%%%%%%%%%%%%%%%%%%%%%%%%%%%%
  $R_{B_s, f_0(980)}$    &  $0.980$  &  $1.000$    &   $1.004$\\ \hline
  %%%%%%%%%%%%%%%%%%%%%%%%%%%%%%
  $R_{B_c, D_s}$           &  $0.981$  &  $1.001$   & $1.006$\\ \hline
  %%%%%%%%%%%%%%%%%%%%%%%%%%%%%%
  $R_{B_s, \phi}$          &  $0.937$  &  $0.998$   &  $0.998$ \\
  $R_{B_s, \phi}^L $       &  $0.991$  &  $1.001$   & $0.999$\\
  $R_{B_s, \phi}^T $       &  $0.902$  &  $0.985$   & $0.997$ \\    \hline
  %%%%%%%%%%%%%%%%%%%%%%%%%%%%%%
  $R_{B_c, D^*_s}$         &  $0.917$  &  $0.995$   &  $0.997$ \\
  $R_{B_c, D^*_s}^L$       &  $0.978$  &  $0.997$   &  $0.997$\\
  $R_{B_c, D^*_s}^T$       &  $0.908$  &  $0.990$   &  $0.997$ \\     \hline
  %%%%%%%%%%%%%%%%%%%%%%%%%%%%%%
  $R_{B,   K^*_2}$         &  $0.932$  &
  $0.996$ &  $0.997$\\
  $R_{B,   K^*_2}^L$       &  $0.971$  &
  $0.998$  &   $0.998$\\
  $R_{B,   K^*_2}^T$       &  $0.902$  &
  $0.985$  &  $0.997$\\   \hline
  %%%%%%%%%%%%%%%%%%%%%%%%%%%%%%
  $R_{B_s,   f_2}$       &  $0.930$  &  $0.995$   &  $0.998$ \\
  $R_{B_s,   f_2}^L$     &  $0.971$  &  $0.998$   & $0.998$ \\
  $R_{B_s,   f_2}^T$     &  $0.902$  &  $0.985$   &  $0.997$\\ \hline
  %%%%%%%%%%%%%%%%%%%%%%%%%%%%%%
  $R_{B,   K_1(1270)}$         &  $0.950$  &
  $1.015$ &  $0.998$\\
  $R_{B,   K_1(1270)}^L$       &  $1.064$  &
  $1.039$  &   $0.999$\\
  $R_{B,   K_1(1270)}^T$       &  $0.901$  &
  $0.985$  &  $0.997$\\   \hline
  %%%%%%%%%%%%%%%%%%%%%%%%%%%%%%
 \end{tabular}
 \end{center}
 \end{table}

A few remarks are given in order.

\begin{itemize}

\item From the decay widths  for $B\to K^*\ell^+\ell^-$, we can see that  in the transverse polarization,  the contribution from $O_7$  is enhanced at low $q^2$,   and thus the $R_{B,M}^T$ is less sensitive to the NP in $O_{9,10}$.
Measurements of the $\mu$-to-$e$  ratio  in  the transverse polarization of $B\to V\ell^+\ell^-$    at low $q^2$   can tell whether the NP is from  the $q^2$ independent contribution in $C_{9,10}$ or the $q^2$ dependent contribution in $C_{7}$.

\item In the central $q^2$  region,  the operators $O_7$ and $O_{9,10}$ will contribute destructively to the transverse polarization of $B\to V\ell^+\ell^-$. Reducing  $C_{9}$ with $\delta C_{9}^\mu<0$  will affect the cancellation, and as a result the decay width for the muonic decay mode will be enhanced. Thus instead of having a ratio smaller than 1, one will obtain a surplus for this ratio.

\item Results for $\Lambda_b\to \Lambda$ with different polarizations are  similar, but it should be pointed out that  differential decay widths in Eq.~\eqref{eq:Lambda_width} have neglected the kinematic lepton mass corrections. Thus the results in the  low $q^2$ region are not   accurate.

\item For the $B\to K_{0,2}(1430)\ell^+\ell^-$ and $B_c\to D_{s}^*$,  the high $q^2$ region has a limited kinematics, and thus the results are difficult  to  be measured.

\item Among the decay processes involved in Table II, a few of them have been experimentally investigated: the branching fractions of $B_s\to \phi\ell^+\ell^-$~\cite{Aaij:2013aln,Aaij:2015esa}, $\Lambda_b\to \Lambda \ell^+\ell^-$~\cite{Aaij:2015xza} and $B_s\to f_0(980)\ell^+\ell^-$~\cite{Aaij:2014lba} have been measured. So for these channels, the measurement of the $\mu$-to-e ratio will be straightforward when enough statistical luminosity is accumulated.

For the other channels, we believe most of them except the $B_c$ decay might also be experimentally measurable, especially at the Belle-II with the designed $50 ab^{-1}$ data and the high luminosity upgrade of LHC.

\item
In FIG.~\ref{fig:NP}, a new particle like  $Z'$ or   leptoquark  can
contribute to the $R_K$ and $R_{K^*}$.   The coupling  strength is
unknown, and in principle  it could be different from the CKM
pattern.  In the SM, the  $B\to \pi\ell^+\ell^-$ and $B_s\to
K\ell^+\ell^-$ have smaller CKM matrix elements. Thus if the NP
contributions had  the same magnitude as in $b\to s\ell^+\ell^-$, their impact in  $B\to
\pi\ell^+\ell^-$ and $B_s\to K\ell^+\ell^-$  would be much larger.
But in many frameworks,  the new physics in $b\to d\ell^+\ell^-$ is suppressed compared to those in $b\to s\ell^-\ell^-$, for recent discussions see Ref.~\cite{Bhattacharya:2014wla}.  This can be   resolved  by experiments in the future.

\item The weak phases from $Z'$ and leptoquark can be different from that in $b\to s\mu^+\mu^-$ or  $b\to d\mu^+\mu^-$, which  may  induce direct CP violations. In the $b\to d\mu^+\mu^-$ process,  the current data on $B\to \pi\mu^+\mu^-$ contains a large uncertainty~\cite{Aaij:2015nea}
\begin{eqnarray}
{\cal A}_{CP}(B^\pm \to \pi^\pm \mu^+\mu^-) =(-0.12\pm0.12\pm0.01).
\end{eqnarray}
This can be certainly refined  in the future.
It should be noticed  that the SM contribution may  also contain CP violation source~\cite{Ali:2013zfa,Li:2015cta} since the up-type quark loop contributions are sizable.
\end{itemize}

%

%%%%%%%%%%%%%%%%%%%%%%%
\section{Conclusions}
\label{sec:conclusions}
%%%%%%%%%%%%%%%%%%%%%%%

Due to the small  branching fractions in the SM, rare decays  of
heavy mesons can  provide  a rich laboratory to search for effects
of physics beyond the SM.  Up to date, quite a few  quantities   in
$B$ decays  have exhibited moderate deviations from the SM. This
happens in both   tree operator and penguin operator induced
processes. The so-called $R_{D(D^*)}$ anomaly gives a hint that the
tau lepton might have a different interaction   with the
light leptons. The  $V_{ub}$ and $V_{cb}$ puzzles   refer  to the
difference for the CKM matrix elements  extracted from the exclusive
and inclusive decay modes. In the $b\to s\ell^+\ell^-$ mode,   the
$P_5'$  in $B\to K^*\ell^+\ell^-$ has received considerable
attentions on both the reliable estimates of hadronic uncertainties
and new physics effects. In addition, LHCb also observed a
systematic deficit with respect to   SM predictions for the
branching ratios of several decay modes, such as $B_s \to \phi
\mu^+\mu^-$~\cite{Aaij:2013aln,Aaij:2015esa}.   Though the statistical significance is low, all these anomalies
 indicate that the NP particles could  be detected  in flavor physics.  %The experimental
%prospect is very promising due to not only the LHC but also the
%forthcoming  Belle-II experiment.

In this work, we have presented an analysis of the recently observed  $R_K$ and $R_{K^*}$ anomalies.   In terms of  the effective operators,  we have performed  a model-independent fit to the  $R_K$ and $R_{K^*}$ data. In the analysis, we have used two sets of form factors and found the results are rather stable against these  hadronic inputs.
Since the statistical significance in $R_K$ and $R_{K^*}$ is rather low,
we  proposed to study   a number of related rare $B, B_s, B_c$ and $\Lambda_b$ decay channels,  and in particular we  have pointed out  that the  $\mu$-to-$e$ ratios of decay widths with different polarizations of  the final state particles,  and in the $b\to d\ell^+\ell^-$ processes are likely more sensitive to the structure of the underlying new physics. After taking into account the new physics contributions, we made theoretical  predictions on  lepton flavor non-universality in these processes which can  stringently   examined by experiments in future.

% %%%%%%%%%%%%%%%%%%%%%%%%%%%%%%%%%
\section*{Acknowledgements}
We thank Yun Jiang and Yu-Ming Wang for useful discussions.
This work is supported  in part   by National  Natural
Science Foundation of China under Grant
 No.11575110, 11655002, 11735010,  Natural  Science Foundation of Shanghai under Grant  No.
15DZ2272100 and No. 15ZR1423100,  by the Young Thousand Talents Plan,   and  by    Key Laboratory for Particle Physics, Astrophysics and Cosmology, Ministry of Education.
% %%%%%%%%%%%%%%%%%%%%%%%%%%%%%%%%%

\appendix

\section{Definitions of $R^{L,T}$ and $R^{0,1}$}

For $B$ decays to vector final state, we define the longitudinal and
transverse ratios $R^L$ and $R^T$ as
\begin{eqnarray}
R^{L,T}_{V}[q_{\rm min}^2,q_{\rm max}^2]\equiv \frac{\int_{q_{\rm
min}^2}^{q_{\rm max}^2}dq^2 d\Gamma^{L,T}(B \to V \mu^+
\mu^-)/dq^2}{\int_{q_{\rm min}^2}^{q_{\rm max}^2}dq^2
d\Gamma^{L,T}(B \to V e^+ e^-)/dq^2},
\end{eqnarray}
where the longitudinal and transverse differential widths are
defined by
\begin{eqnarray}
d\Gamma^{L}(B \to V \mu^+ \mu^-)/dq^2&=&\frac34 I^c_1-\frac14
I^c_2,\\
d\Gamma^{T}(B \to V \mu^+ \mu^-)/dq^2&=&\frac32 I^s_1-\frac12 I^s_2,
\end{eqnarray}
$V$ denotes a vector final state. The expressions for $I^{c,s}_1$
and $I^{c,s}_2$ are given by Eq.~\eqref{eq:I1I2}.

For $\Lambda_b\to \Lambda\ell^+\ell^-$ decays, we define ratios with
equal or different polarization as~\cite{Boer:2014kda}
\begin{eqnarray}
  R^{0,1}[q_{\rm min}^2,q_{\rm max}^2]\equiv \frac{\int_{q_{\rm
min}^2}^{q_{\rm max}^2}dq^2 d\Gamma^{0,1}(\Lambda_b \to \Lambda
\mu^+ \mu^-)/dq^2}{\int_{q_{\rm min}^2}^{q_{\rm max}^2}dq^2
d\Gamma^{0,1}(\Lambda_b \to \Lambda e^+ e^-)/dq^2},
\end{eqnarray}
the superscript $0$ means that the $\Lambda_b$ and  $\Lambda$ have
the same polarization, while $1$ corresponds to different
polarizations. The expressions for $d\Gamma^{0,1}/dq^2$ are
\begin{eqnarray}
  d\Gamma^{0}(\Lambda_b \to \Lambda
\mu^+ \mu^-)/dq^2&=& 2 K^0_{1ss},\\
d\Gamma^{1}(\Lambda_b \to \Lambda
\mu^+ \mu^-)/dq^2&=& 2 K^1_{1ss}+K^1_{1cc},\\
\end{eqnarray}
$K^{0,1}_{1ss}$ and $K^{1}_{1cc}$ are defined by
\begin{eqnarray}
  K^0_{1ss}&=&\frac12(|A^R_{\perp 0}|^2+|A^R_{\| 0}|^2+|A^L_{\perp 0}|^2+|A^L_{\|
  0}|^2),\\
  K^1_{1ss}&=&\frac14(|A^R_{\perp 1}|^2+|A^R_{\| 1}|^2+|A^L_{\perp 1}|^2+|A^L_{\|
  1}|^2),\\
  K^1_{1cc}&=&\frac12(|A^R_{\perp 1}|^2+|A^R_{\| 1}|^2+|A^L_{\perp 1}|^2+|A^L_{\|
  1}|^2).
\end{eqnarray}
The $A$ functions have already been defined in
Eq.~\eqref{eq:Afunction}.

%%%%%%%%%%%%%%%%%%%%%%%%%%%%%%%%%%%%%%%%%%%%%%%%%%%%%%%%%%

\end{document}